\documentclass[useAMS,usenatbib]{mn2e}
\pdfoutput=1
\usepackage{amssymb}
\usepackage{amsfonts}
\usepackage{natbib}
\usepackage{epsfig}
\usepackage{mncite}

\newcommand{\msun}{M_{\sun}}

\newcommand{\msigma}{M_{\rm BH}-\sigma}

\begin{document}

\title[$\msigma$ relation]{Journey to the $M_{\rm BH} - \sigma$ relation: the fate of 
low mass black holes in the Universe}

\author[Volonteri \&  Natarajan] {Marta Volonteri$^{1}$ and Priyamvada 
Natarajan$^{2,3}$\\
$^{1}$ Department of Astronomy, University of Michigan, Ann Arbor, MI, USA\\
$^2$ Department of Astronomy, Yale University, P. O. Box 208101, 
New Haven, CT 06511-208101, USA \\
$^3$ Radcliffe Institute for Advanced Study, Harvard University, 10 Garden
Street, Cambridge, MA 02138, USA \\
}

\maketitle

\begin{abstract}
In this paper, we explore the establishment and evolution of the
empirical correlation between black hole mass ($M_{\rm BH}$) and
velocity dispersion ($\sigma$) with redshift.  We trace the growth and
accretion history of massive black holes (MBHs) starting from high
redshift seeds that are planted via physically motivated
prescriptions. Two seeding models are explored in this work: `light
seeds', derived from Population III remnants, and `heavy seeds',
derived from direct gas collapse. Even though the seeds themselves do not
satisfy the $M_{\rm BH} - \sigma$ relation initially, we find that
the relation can be established and maintained at all times if
self-regulating accretion episodes are associated with major
mergers. The massive end of the $M_{\rm BH} - \sigma$ relation is
established early, and lower mass MBHs migrate onto it as hierarchical
merging proceeds. How MBHs migrate toward the relation depends
critically on the seeding prescription. Light seeds initially lie well
below the $M_{\rm BH} - \sigma$ relation, and MBHs can grow via steady
accretion episodes unhindered by self-regulation. In contrast, for the
heavy seeding model, MBHs are initially over-massive compared to the
empirical correlation, and the host haloes assemble prior to
kick-starting the growth of the MBH.  We find that the existence of
the $M_{\rm BH} - \sigma$ correlation is purely a reflection of the
merging hierarchy of massive dark matter haloes. The slope and scatter
of the relation however appear to be a consequence of the seeding
mechanism and the self-regulation prescription. We expect flux
limited AGN surveys to select MBHs that have already migrated onto the
$\msigma$ relation. Similarly, LISA is also likely to be biased toward
detecting merging MBHs that preferentially inhabit the $M_{\rm BH}
-\sigma$. These results are a consequence of major mergers
being more common at high redshift for the most massive, biased,
galaxies that host MBHs which have already migrated onto the $\msigma$
relation. We also predict the existence of a large population of low
mass `hidden' MBHs at high redshift which can easily escape detection.
Additionally, we find that if MBH seeds are massive, $\sim
10^5\,M_{\odot}$, the low-mass end of the $\msigma$ flattens towards
an asymptotic value, creating a characteristic `plume'.
\end{abstract}

\section{Introduction}

The demography of local galaxies suggests the that almost every galaxy
hosts a quiescent super-massive black hole (MBH) at the present time and
the properties of the MBH are correlated with those of the host. In
particular, recent observational evidence points to the existence of a
strong correlation between the mass of the central MBH and the
velocity dispersion of the host spheroid (Tremaine et al. 2002;
Ferrarese \& Merritt 2001, Gebhardt et al. 2002; Marconi \& Hunt 2003;
H\'aring \& Rix 2004; G\"ultekin et al. 2009) and possibly the host
halo (Ferrarese 2002) in nearby galaxies. It is currently unclear if
these correlations hold at higher redshift, or if the scalings evolve
with cosmic time. These correlations strongly suggest co-eval growth of
the MBH and the stellar component via likely regulation of the gas
supply in galactic nuclei (Silk \& Rees 1998; Kauffmann \& Haehnelt
2000; Fabian 2002; King 2003; Thompson, Quataert \& Murray 2005;
Natarajan \& Treister 2009).

The current phenomenological approach to understanding the assembly of
MBHs involves data from both high and low redshifts.  These data are
used to construct a consistent picture that is in consonance with the
larger framework of the growth and evolution of structure in the
Universe (for example: Haehnelt, Natarajan \& Rees 1998; Haiman \&
Loeb 1998; Kauffmann \& Haehnelt 2000; 2002; Wyithe \& Loeb 2002;
Volonteri et al. 2003; Di Matteo et al. 2003; Steed \& Weinberg
2004). Major mergers appear to drive the establishment of the correlations 
between MBH masses and their host properties (Robertson et al. 2006; Peng 2007;
Hopkins et al. 2007a,b) and it also appears that these correlations
are possibly linear projections of a more universal MBH fundamental
plane relation (Hopkins et al. 2007a). The observed correlations offer
insight into how the dynamics of the merger process establish these
relations. 

In a companion paper (Volonteri, Lodato \& Natarajan 2007) we explored 
the evolution of MBHs with cosmic history starting from physically 
motivated MBH formation models.  We investigated the observational signatures 
by following the mass assembly of these black hole seeds to the present time. 
We showed that the low-redshift population evolved from physically motivated 
seeds agrees nicely with current constraints  (mass function of MBHs at $z=0$; 
the integrated mass density of black holes and the luminosity function of AGN as a
function of redshift).

In this paper, we address the establishment of the correlation 
between MBH masses and the velocity dispersion of their host, 
by focusing on two relevant questions as we track
the journey of black holes onto the observed $z = 0$ $M_{\rm
BH}-\sigma$ relation, (i) are the correlations established independently
of galaxy mass, and (ii) can observations at $z>0$ select
samples unbiased with respect to the $M_{\rm BH} - \sigma$ relation.

The structure of our paper is as follows: in the first and second
sections we outline very briefly the basic methodology adopted to
track the merger history, in the third section, we focus on the
details of the $M_{\rm BH} - \sigma$ relation and its establishment
with epoch and mass (in Section 4). The observational consequences of our model are
described in Section 5 and our conclusions are discussed in the final
section of this paper.

\section{Methodology}

We ground our models in the framework of the standard paradigm for the
growth of structure in a $\Lambda$CDM Universe---a model that has
independent validation, most recently from {\it Wilkinson Microwave
Anisotropy Probe (WMAP)} measurements of the anisotropies in the
cosmic microwave background (Spergel et al.\ 2003; Page et al.\
2003). Structure formation is tracked in cosmic time by keeping a
census of the number of collapsed dark matter haloes of a given mass
that form; these provide the sites for harboring MBHs. The computation
of the mass function of dark matter haloes is done using the extended
Press-Schechter theory (Lacey \& Cole 1993) and Monte-Carlo
realizations of merger trees (Volonteri et al. 2003). Monte-Carlo
merger trees are created for present day haloes and propagated back in
time to a redshift of $\sim$ 20. With the merging history thus
determined, the haloes are then populated with seed MBHs. The halo
merger sequence is followed and black holes are grown embedded in
their dark matter halo.  

\subsection{The initial BH seeding model}

We compare two distinct types of seeds: `light seeds', derived from
Population III remnants, and `heavy seeds', where we plant the initial
seeds in the dark matter haloes according to the prescription
described in Volonteri, Lodato \& Natarajan (2007) as per the
physically motivated model developed by Lodato \& Natarajan (2007;
2006).

In the `heavy seeds' scenario, massive seeds with $M\approx
10^5-10^6M_{\odot}$ can form at high redshift ($z>15$), when the
intergalactic medium has not been significantly enriched by metals
\citep{koushiappas04,begelman06,LN06,LN07}.  Here we refer to
\citet{LN06, LN07}, for more details of the seeding model, wherein the
development of non-axisymmetric spiral structures drives mass infall
and accumulation in a pre-galactic disc with primordial
composition. The mass accumulated in the center of the halo (which
provides an upper limit to the MBH seed mass) is given by:
\begin{equation}
M_{\rm BH}= m_{\rm d}M_{\rm halo}\left[1-\sqrt{\frac{8\lambda}{m_{\rm d}Q_{\rm c}}\left(\frac{j_{\rm d}}{m_{\rm d}}\right)\left(\frac{T_{\rm gas}}{T_{\rm vir}}\right)^{1/2}}\right] 
\label{mbh}
\end{equation}
for 
\begin{equation}
\lambda<\lambda_{\rm max}=m_{\rm d}Q_{\rm c}/8(m_{\rm d}/j_{\rm d}) (T_{\rm
  vir}/T_{\rm gas})^{1/2}
\label{lambdamax} 
\end{equation}
and $M_{\rm BH}=0$ otherwise. Here $\lambda_{\rm max}$ is the maximum
halo spin parameter for which the disc is gravitationally unstable,
$m_d$ is the gas fraction that participates in the infall and $Q_{\rm
c}$ is the Toomre parameter.  The efficiency of MBH formation is
strongly dependent on the Toomre parameter $Q_{\rm c}$, which sets the
frequency of formation, and consequently the number density of MBH
seeds. Guided by our earlier investigation, we set $Q_{\rm c}=2$ (the
intermediate efficiency model) as described in Volonteri, Lodato \&
Natarajan (2007).

The efficiency of the seed assembly process ceases at large halo
masses, where the disc undergoes fragmentation instead. This occurs
when the virial temperature exceeds a critical value $T_{\rm max}$,
given by:
\begin{equation}
\frac{T_{\rm max}}{T_{\rm gas}}=\left(\frac{4\alpha_{\rm c}}{m_{\rm
d}}\frac{1}{1+M_{\rm BH}/m_{\rm d}M_{\rm halo}}\right)^{2/3},
\label{frag}
\end{equation}
where $\alpha_{\rm c}\approx 0.06$ is a dimensionless parameter measuring the
critical gravitational torque above which the disc fragments \citep{RLA05}.

To summarize, every dark matter halo is characterized by its mass $M$
(or virial temperature $T_{\rm vir}$) and by its spin parameter
$\lambda$. The gas has a temperature $T_{\rm gas}=5000$K. If
$\lambda<\lambda_{\rm max}$ (see eqn.~\ref{lambdamax}) and $T_{\rm
vir}<T_{\rm max}$ (eqn.~\ref{frag}), then we assume that a seed BH of
mass $M_{\rm BH}$ given by eqn.~(\ref{mbh}) forms in the center. The
remaining relevant parameters are $m_{\rm d}=j_{\rm d}=0.05$,
$\alpha_{\rm c}=0.06$ and here we consider the $Q_{\rm c}=2$ case.


\begin{figure} 
\includegraphics[width=\columnwidth]{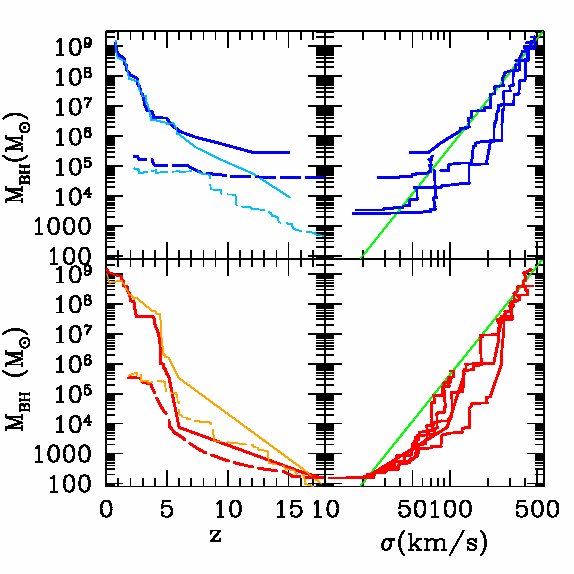}
  \caption{Tracks of MBH growth as a function of redshift  or velocity dispersion along the history of 
  a $4\times10^{13}\,M_{\odot}$ halo. 
  Top: `heavy seeds'; bottom: 'light seeds'.
  When we track the MBH growth as a function of redshift, we show with a solid curve the MBH in the main halo; with 
  a dashed curve a MBH in a satellite galaxy. The thick lines show growth histories extracted from our models, the thin lines 
  show the mass the MBH would have if it sat on the $\msigma$ relation at the
  times when we record MBH masses.  If seeds are light, the MBHs typically have to catch up with their  host, viceversa if seeds are 
   heavy their growth is impeded if feedback effects that limit the MBH mass
  are at work.  }
   \label{fig0}
\end{figure}

In the `heavy seed' model, MBHs form (i) only in haloes within a narrow range of virial
temperatures ($10^4$ K$ < T_{vir}<1.4\times10^4$ K), hence, halo
velocity dispersion ($\sigma \simeq 15\,{\rm km\,s}^{-1} $), and (ii)
for each virial temperature all seed masses below $m_d\,M$ modulo the
spin parameter of the halo are allowed (see equations~1 and 3). The
seed mass function peaks at $10^5\msun$, with a steep drop at
$3\times10^6\msun$.  We refer the reader to Lodato \& Natarajan (2007)
and Volonteri, Lodato \& Natarajan (2008) for a discussion of the mass
function (and related plots).  Here we stress that given points () and (ii) above, 
the initial seeds do not satisfy the local $M_{\rm BH} -\sigma$ relation, in fact the seed masses are not
correlated with $\sigma$, as shown in the lower left hand panels of
Fig.~1 (see the almost vertical line in the $z=4$ panels).

In the Population III remnants model (`light seeds'), MBHs form as end-product of the very first 
generation of stars, with masses $m_{\rm seed}\sim$ few$\times10^2\msun$.   
The first stars are believed to form at $z\sim 20-30$ in halos which represent high-$\sigma$ 
peaks of the primordial density field.  The main coolant, in absence of metals, is molecular 
hydrogen, which is a rather inefficient coolant.  The inefficient cooling might lead to a very 
top-heavy initial stellar mass function, and in particular to the production of  an early generation 
of  very massive stars (Carr, Bond, \& Arnett 1984). If stars form above 260 $M_\odot$,  they 
would rapidly collapse to massive black holes with little mass loss (Fryer, Woosley, \& Heger 2001), 
i.e., leaving behind seed MBHs with masses $M_{BH} \sim 10^2-10^3\,M_\odot$ 
(Madau \& Rees 2001).  

The main features of a scenario for the hierarchical assembly of MBHs left over by 
the first stars in a $\Lambda$CDM cosmology
have been discussed by Volonteri, Haardt, \& Madau (2003) and Volonteri \& Rees 2006.   
Stars, and their remnant MBHs, form in isolation within mini-halos above the cosmological Jeans mass 
collapsing at $z \geqslant 20$ from rare $\nu$-$\sigma$ peaks of the primordial density 
field (Madau \& Rees 2001). We here consider $\nu=$3.5, that is, very rare peaks of the primordial density field
\citep{volonteri03}. We assume that seeds form in the mass range $125<M_{\rm BH} <1000\,\msun$, from an initial stellar mass function
with slope $-2.8$. Population III remnants do not satisfy any $M_{\rm
BH} -\sigma$ relation either, as shown in Fig.~2 (lower left hand
panels).

When a halo enters the merger tree we assign seed MBHs by determining
if the halo meets all the requirements described above (separately for each model). As we do not
trace the metal enrichment of the intergalactic medium
self-consistently, we consider here a sharp transition threshold, and
assume that MBH seed formation ceases at $z\approx 15$ (cfr. Volonteri
et al. 2008).

\section{Tracking the growth of MBHs}

We follow the evolution of the MBH population resulting from the seed
formation processes briefly outlined above in a $\Lambda$CDM
Universe. We simulate the merger history of 2 sets of present-day
haloes, one with mass $2\times10^{12}\,\msun$ mimicking the Milky
Way (MW) and the other with mass $4\times10^{13}\,\msun$ mimicking a
massive elliptical (ET), via a Monte-Carlo algorithm based on the
extended Press-Schechter formalism. 

Here and throughout the paper we use the
velocity dispersion of the halo as a proxy for the central velocity
dispersion $\sigma$ (Ferrarese et al. 2002, Pizzella et
al. 2005). Every halo entering the merger tree is assigned a spin
parameter drawn from the lognormal distribution in $\lambda_{\rm
spin}$ found in numerical simulations, with mean $\bar \lambda_{\rm
spin}=0.05$ and standard deviation $\sigma_\lambda=0.5$ (e.g.,
\citealt{warren92,cole96,bullock01,bosch02}). We assume that the spin
parameter of a halo is not modified by its merger history, as no
consensus exists on this issue at the present time.

\begin{figure*} 
\includegraphics[width=\columnwidth]{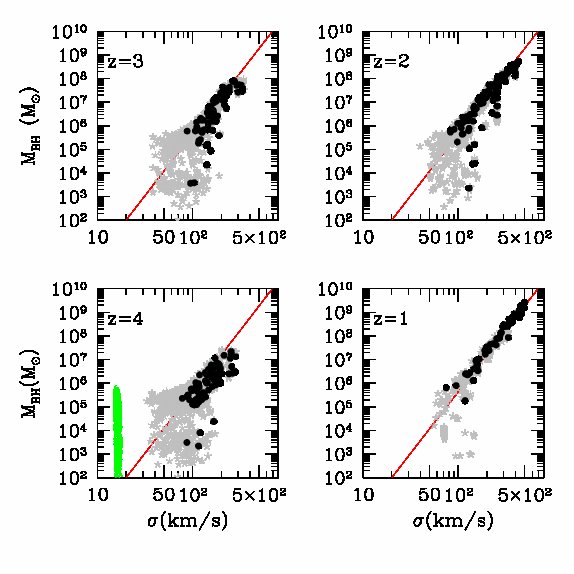}
\includegraphics[width=\columnwidth]{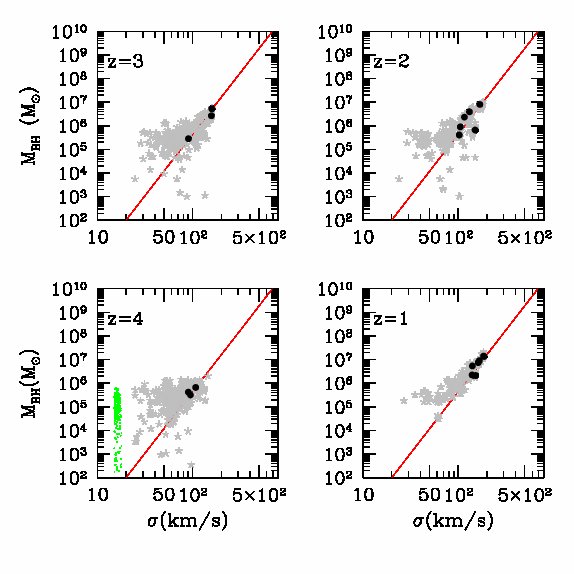}
  \caption{The $\msigma$ relation for MBHs at different redshifts
  along the merging history of a $4\times10^{13}\,M_{\odot}$ halo (ET,
  left), and for a $2\times10^{12}\,M_{\odot}$ halo (MW, right). The
  sample above comprises 20 realisations for each halo mass, and for
  each halo we include all the progenitors that exist at a given
  cosmic time.  MBHs evolve from an initial population of seeds based
  on the model by Lodato \& Natarajan (2006), with $Q_c=2$ (the lack
  of any initial $\msigma$ correlation for seeds is clearly seen seen
  in the far left corner of the $z=4$ panels, green points).  Note
  that all the initial seeds in this model are over-massive compared
  to the local $\msigma$ relation. Grey points: all central MBHs in
  the progenitors of the galaxy at the specified redshift. Black
  points and triangles: all systems experiencing a MBH-MBH merger
  within the same redshift range (triangles indicate the less massive
  MBH of the pair). The velocity dispersion plotted is that of the
  merger remnant. Note the `plume' of MBHs at $\sigma
  <\,50\,\rm{km\,s}^{-1}$ that clearly persists even at $z = 2$ from
  the earliest epochs.}
   \label{fig1}
\end{figure*}

\begin{figure*}   
  \includegraphics[width=\columnwidth]{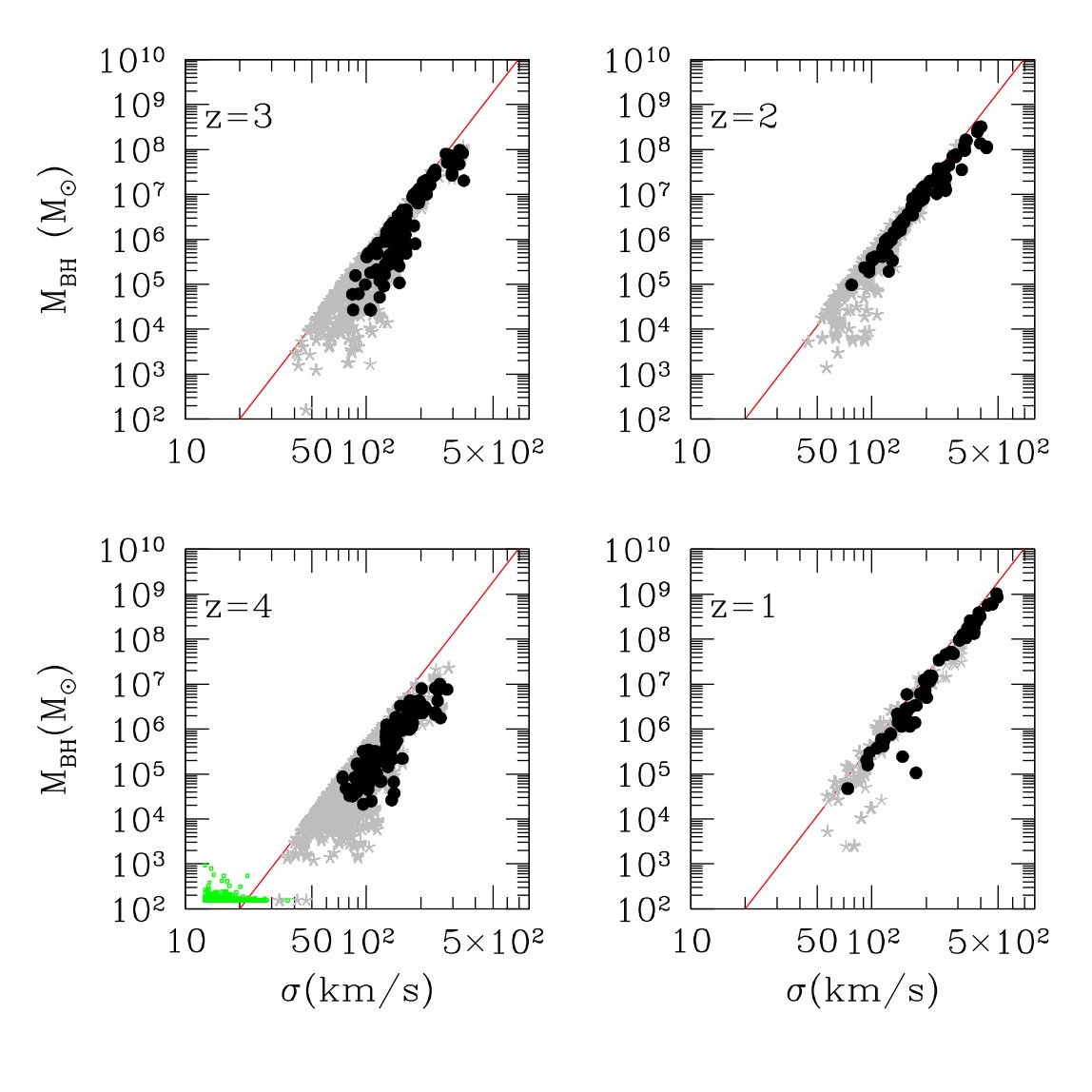}
  \includegraphics[width=\columnwidth]{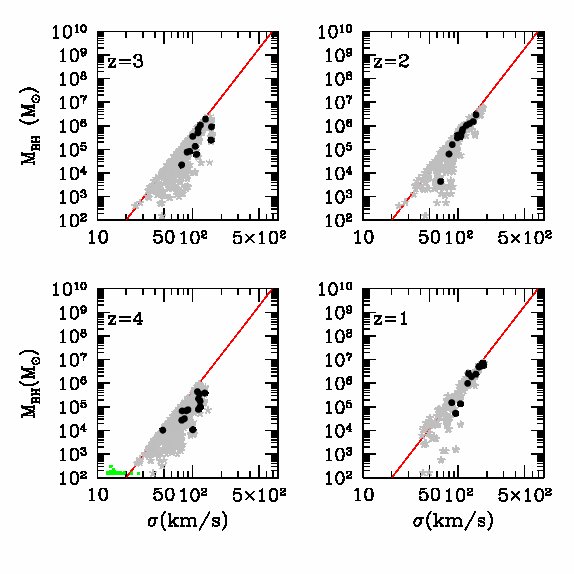}
  \caption{Same as Fig.~\ref{fig1}, for Population III remnant
seeds. The lack of an initial $\msigma$ correlation for these seeds is
also evident here and is shown at the bottom of the $z=4$ panels for
the MW and ET halo realizations (green points). Note that the sharp
difference in the assignment of the initial seed population $z = 15$
in the two models is evident even in the $z = 4$ panel.}
   \label{fig2}
\end{figure*}

We assume that, after seed formation ceases, the population of MBH
progenitors evolves according to a `merger driven scenario', as
described in Volonteri et al. (2003; 2006). An accretion episode is
assumed to occur as a consequence of every major merger (mass ratio
larger than 1:10) event. Each MBH accretes an amount of mass, 
$\Delta M=9\times 10^7\msun(\sigma/200{\rm km\,s^{-1}})^4$, that corresponds to 90\% of the $M_{\rm
BH}-\sigma_*$ relation of its host halo (Ferrarese et al. 2002). 
This choice allows us to take into account the contribution of mergers. 
If a MBH increases its mass beyond the $\msigma$ relation, we shut its growth. 
During this phase a MBH would be classified as an AGN.
The rate at which mass is accreted scales with the Eddington rate for the MBH, and is based on the
results of galaxy merger simulations, which also heuristically track
accretion onto a central MBH (di Matteo et al.  2005; Hopkins et
al. 2005a). We impose a lower limit to the Eddington ratio of $f_{\rm
Edd}=10^{-3}$. Accretion starts after a
dynamical timescale and lasts until the MBH has accreted $\Delta M$.
The lifetime of AGN therefore depends on how much mass it accretes in each episode:
\begin{equation}
t_{\rm AGN}=t_{Edd}\,f_{\rm Edd} \frac{\epsilon}{1-\epsilon}\ln(M_{\rm fin}/M_{\rm BH});
\end{equation}
where $M_{\rm fin}=\min(M_{\rm BH}+\Delta M,\msigma)$;
$\epsilon$ is the radiative efficiency (which depends with the MBH spin, $\langle \epsilon \rangle \approx 0.2$, 
assuming coherent accretion, Berti \& Volonteri 2008)
and $t_{Edd}=0.45$ Gyr. The farther away a MBH is from the $\msigma$, the longer it shines before 
accretion is shut when it reaches the $\msigma$ limit. 

In this scheme we assume that  MBHs
accrete a gas mass that scales with the fifth power of the circular
velocity (or equivalently $\sigma$) of the host halo. \textit{We do not assume any evolution of either slope or
normalisation of this scaling with redshift.}  Given this assumption,
it is clearly not our goal to study the evolution of the slope and
normalisation of the observed $\msigma$ relation or the scatter with
redshift. We focus on analysing how MBH seeds, that do not initially
satisfy any correlation with the host mass or velocity dispersion,
migrate towards the observed correlation at $z = 0$ as a function of
cosmic time. In this context, {\it the exact scaling of the accreted
mass does not affect our results, as long as accretion is
merger-driven and it establishes a clear correlation between hole and
host}.

In a hierarchical universe, where galaxies grow by mergers, MBH
coalescences are a natural consequence, and we trace their
contribution to the evolving MBH population (cfr. Sesana et al. 2007
for details on the dynamical modeling). During the final phases of a
MBH merger, emission of gravitational radiation drives the orbital
decay of the binary. Recent numerical relativity simulations suggest
that merging MBH binaries might be subject to a large `gravitational
recoil': a general-relativistic effect due to the non-zero net linear
momentum carried away by gravitational waves in the coalescence of two
unequal MBHs (Fitchett 1982; Redmount \& Rees 1992). Radiation recoil
is a strong field effect that depends on the lack of symmetry in the
system, and for merging MBHs with high spin in particular orbital
configurations, the recoil velocity can be as high as a few thousands
of kilometers per second. We include the effects of gravitational
recoil by adopting the fitting formula proposed by Lousto \& Zlochower
(2008, see also Baker et al. 2008).  MBHs that are displaced from
galaxy centres by the gravitational recoil effect produce a population
of {\it wandering} MBHs and AGNs as explored in earlier work
(Volonteri \& Perna 2005, Volonteri \& Madau 2008, Devecchi et
al. 2008).

\section{Tracking the $M_{\rm BH} - \sigma$ relation}

We present the results of tracking the assembly history of MBHs in 2
classes of galaxies, (i) a dark matter halo with mass
$2\times10^{12}\,M_{\odot}$ that hosts a MW type galaxy and (ii) a
more massive dark matter halo, $4\times10^{13}\,M_{\odot}$, that hosts
a massive early type (ET) galaxy. The progenitors of the MBHs in each of these host
haloes are tracked and plotted as measured at a given epoch. We
analyse 20 realisations for each halo, to account for cosmic variance.
Examples of growth histories are shown in
Fig.~\ref{fig0}, while statistical $\msigma$ relations are shown in Fig~\ref{fig1}  
and Fig.~\ref{fig2} for the two seed models.

As outlined earlier, in propagating the seeds it is assumed that
accretion episodes and therefore growth spurts are triggered only by
major mergers. We find that in a merger-driven scenario for MBH growth
the most biased galaxies at every epoch host the most massive MBHs
that are most likely already sitting on the $M_{\rm BH} -\sigma$
relation.  Lower mass MBHs (below $10^6\,\msun$) are instead off the
relation at $z = 4$ and even at $z = 2$. These baseline results are
{\it independent of the seeding mechanism}.  
In the `heavy seeds' scenario, most of the MBH seeds start out {\it
well above} the $z=0$ $\msigma$, that is, they are `overmassive'
compared to the local relation.  Seeds form only in haloes within a
narrow range of velocity dispersion ($\sigma \simeq 15\,{\rm
km\,s}^{-1} $, see equations~1 and 3, and Fig.~1).  The MBH mass
corresponding to $\sigma \simeq 15\,{\rm km\,s}^{-1}$, according to
the local $M_{\rm BH} -\sigma$ relation, would be $\sim 3\times 10^3
\msun$.  The mass function instead peaks at $10^5\msun$ (Lodato \&
Natarajan 2007).  As time elapses, all haloes are bound to grow in
mass by mergers.  The lowest mass haloes, though, experience mostly
minor mergers, that do not trigger accretion episodes, and hence do
not grow the MBH. The evolution of these systems can be described by
a shift towards the right of the $\msigma$ relation: $\sigma$
increases, but $M_{\rm BH}$ stays roughly constant. Such systems are
clearly seen at $z=1$ in Fig.~\ref{fig1}, with $M_{\rm
BH}\sim10^5\msun$ and $\sigma<100\,{\rm km\,s}^{-1}$.  Effectively,
for the lowest mass haloes growth of the galaxy and the central MBH
are not coeval but rather sequential.

In the case of Population III seeds as well there is initially no
correlation between seed mass and halo mass or velocity
dispersion. Here we have assumed that the seeds form in the mass range
$125<M_{\rm BH} <1000\,\msun$.  The initial $\msigma$ relation would
therefore appear as a horizontal line at $\sim 200\msun$ (shown at the
bottom of Fig.~\ref{fig2}, $z=4$ panels). In this case MBHs migrate
onto the $\msigma$ always from {\it below}, as seeds are initially
`undermassive' compared to the local relation (Fig.~\ref{fig0}, bottom panels). Underfed survivors of
the seed epoch shift towards the right of the $\msigma$ relation and
lie in the lower left corner of Fig.~\ref{fig2}, with $M_{\rm
BH}\sim10^2-10^3\msun$ and $\sigma<100\,{\rm km\,s}^{-1}$.

There appears to be a distinct difference between the journey of MBH
seeds onto the $\msigma$ relation predicted by the two seeding models
considered here. The Population III seeds start life `undermassive'
lying initially below the local $\msigma$ and they transit up to the
relation by essentially growing the MBH without significantly altering
$\sigma$. In contrast, the massive seeds start off above the local
$\msigma$ relation, and migrate onto it by initially growing $\sigma$,
after which further major mergers trigger accretion episodes and
therefore growth spurts for the MBHs.  When MBHs are more massive than
expected compared to the $\msigma$ relation, accretion is terminated very rapidly in 
our scheme (physically, we expect feedback to be responsible for shutting
down accretion, see, e.g., Silk \& Rees 1998, Fabian 2002).

\begin{figure}   
  \includegraphics[width=\columnwidth]{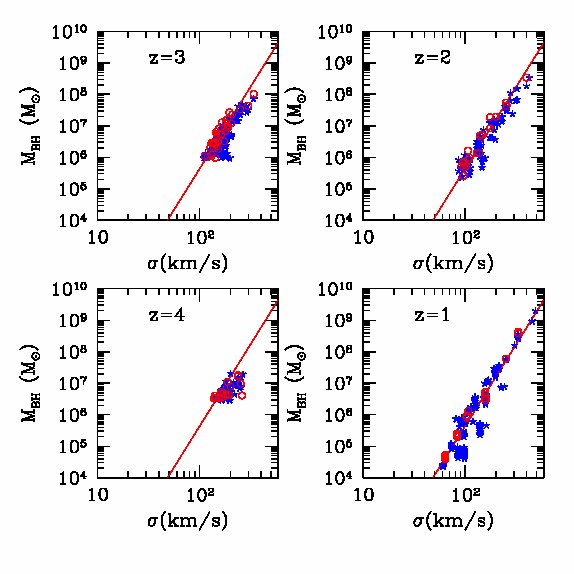}
  \caption{The $\msigma$ relation for active MBHs at different
  redshift slices in the ET progenitors. These MBHs would be observed
  as AGNs.  We imposed a flux threshold, 10$^{-16}$ erg s$^{-1}$ cm$^{-2}$ (bolometric). 
  Stars: Massive seeds based on the model by Lodato \&
  Natarajan (2006), with $Q_c=2$. Circles: seeds based on Population
  III star remnant models (Volonteri et al. 2003).  The figure shows
  both central and satellite MBHs (satellite holes are shown at the
  $\sigma$ of the host halo). The sample comprises all the progenitors
  of 20 $z=0$ haloes.}
   \label{fig3}
\end{figure}

\subsection{What anchors the $\msigma$ relation?}

It appears that major mergers that trigger accretion episodes are what
set up the relation initially at high redshift. Our conclusions in
this regard are in agreement with those reached by alternative
arguments, for instance see Peng (2007) and Robertson et
al. (2006). Biased peaks in the halo mass distribution, which are the
sites for the formation of the largest galaxies, host the earliest
massive MBHs that fall on the relation. Hence, the $\msigma$
correlation is established first for MBHs hosted in the largest haloes
present at any time. MBHs in small galaxies lag behind, as their hosts
are subject to little or no major merger activity. In many cases the
MBHs remain at the original seed mass for billions of years (e.g., see
Fig.~\ref{fig1}, the $z=1$ panel).  We find that these conclusions
hold irrespective of our initial seeding mechanism and the relation
tightens considerably from $z = 4$ to $z = 1$, especially for MBHs
hosted in haloes with $\sigma>100$ $\rm{km\, s}^{-1}$.  We find that
if black hole seeds are massive, $\sim 10^5\,M_{\odot}$, the low-mass
end of the $\msigma$ flattens at low masses towards an asymptotic
value, creating a characteristic `plume'. This `plume' consists of
ungrown seeds, that merely continue to track the peak of the seed mass
function at $M_{\rm BH}\sim10^5\,\msun$ down to late times. For the
Population III seed case, since the initial seed mass is very small,
the plume of MBHs with $M_{\rm BH}\sim10^5-10^6\,\msun$ in haloes with
$\sigma\sim 40-50\,{\rm km\, s^{-1}}$ disappears.


\section{Observational consequences}

We track MBH assembly histories with a view to understanding two kinds
of observations, observations of actively accreting MBHs as probed
by\textit{ flux limited AGN surveys} and potential observations of
\textit{gravitational waves emitted by merging MBHs}. Note that in our
model not every galaxy merger causes a merger of MBHs as one of the
two galaxies might not be seeded. If the halo mass ratio is 1:10 or
higher, every galaxy merger (where at least one of the galaxy hosts a
MBH) triggers accretion and therefore such cases will be detected as
an AGN. AGNs are therefore more common than MBH mergers, in our
scheme.

\subsection{Seed signatures: the AGN population}

 Since it is during accretion episodes  that MBHs move 
onto the $M_{\rm BH} - \sigma$ relation, AGN are better tracers 
of the correlation itself, and worse tracers of the original seeds. 
Differences between seeding models appear only at the low--mass end. We predict the 
existence of many low luminosity accretors with masses off the relation at $z = 4$ down to $z = 3$.
These `outliers' are mostly objects with $M_{\rm BH}<10^6\,\msun$, making them rather faint sources.
For instance, for an Eddington ratio of 0.1, this black hole mass corresponds to an
X--ray luminosity in the [2-10] kev band of  $7.8\times10^{42}$ erg s$^{-1}$,
or B-band luminosity $1.5\times10^{43}$ erg s$^{-1}$. At $z=3$ these luminosities correspond to fluxes of order 
a few times 10$^{-16}$ erg s$^{-1}$ cm$^{-2}$ (as a reference, 
the Chandra Deep Field North has a flux limit $3\times10^{-16}$ erg s$^{-1}$ cm$^{-2}$). 

The population of active MBHs shining above a  flux limit 10$^{-16}$ erg s$^{-1}$ cm$^{-2}$ (bolometric)  
in the history of our ET galaxies is shown in Figure~\ref{fig3}. 
From Figure~\ref{fig3} we note that the seed scenarios are 
less distinguishable for active MBHs than for the case of quiescent MBHs 
(Fig.~\ref{fig1} and Fig.~\ref{fig2}).  The massive end of M-sigma, as traced by AGN, 
is well populated at $z=4$ and  $z=3$  and its only at $z < 2$ that lower masses get on to the
relation. The figure also shows that within the mass range probed by
current flux-limited survey seed formation models are indistinguishable. 
The `outliers' off the $M_{\rm BH} - \sigma$,  with $M_{\rm BH}<10^6\,\msun$,
are currently not easily observable, but future, planned X-ray missions with higher
sensitivity might uncover this population. 


Since MBHs move on to the $\msigma$ relation 
starting from the most massive systems at any time, the implication of our result is that
flux limited AGN surveys tend to be biased toward finding MBHs that
preferentially fall and anchor the $M_{\rm BH} - \sigma$
relation. Flux limited surveys indeed preferentially select the most
massive accreting MBHs residing in the most massive galaxies (Lauer et
al. 2007), assuming that MBHs accrete below the Eddington rate (e.g.,
Kelly et al. 2008).

\subsection{Seed signatures:  MBH mergers and  gravitational waves}

An alternative to AGN observations in electromagnetic bands is the
detection of MBHs via gravitational radiation, that would be
detectable by {\it LISA}. The merger rate of MBHs in our models, and the detectability of 
binaries has been discussed in Sesana et al. (2007), where the impact of different `seed' 
formation scenarios was taken into account.  

Since the focus of this paper are high-redshift objects, we assume that merging is driven by 
dynamical friction, which has been shown to efficiently  drive the MBHs 
in the central regions of the newly formed  galaxy when the mass ratio of the satellite halo to the 
main halo is sufficiently  large, $\geqslant 1:10$ 
and galaxies are gas--rich (Callegari et al. 2008).  The available simulations 
(Escala et al. 2004; Dotti et al. 2006; Mayer et al. 2006) 
show that the binary can shrink to about parsec or slightly subparsec scale by 
dynamical friction against gas. 

We refer the reader to  Sesana et al. (2007) and Sesana et al. (2005)
for a detailed discussion of how we model the gravitational wave emission and the 
expected event rate. Detection of gravitational radiation provides accurate measurements of 
the mass of the components of MBH binaries prior to  merger, and the mass of the single 
merger remnant. Additionally, the mass of `single' MBHs can be determined by the inspiral 
of an extreme or intermediate mass-ratio compact object (EMRI/IMRI, Miller 2005). 
We will discuss EMRI/IMRI events in section 5.3.

When we track the merging population, we find that MBH-MBH mergers
also preferentially sample the region of space
where MBHs lie on the $\msigma$ relation. This is once again a
consequence of halo bias.  Both formation models that we investigate
in this paper require deep potential wells for gas retention and
cooling as a prerequisite for MBH formation.  Haloes where massive
seeds can form are typically 3.5--4 $\sigma$ peaks of the density
fluctuation field at $z>15$, (the host haloes in the direct collapse
model are slightly more biased than in the Population III remnant
case). MBH seeding is therefore infrequent, MBHs are rare and as a
consequence MBH-MBH mergers are events that typically involve only the
most biased haloes at any time. 

In typical mergers we find that the higher mass black hole in the binary
tends to sit on or near the expected $\msigma$ relation for
the host (which corresponds to the newly formed galaxy after the merger).  
The mass of the secondary generally provides clues to the
dynamics of the merger, rather than to the $\msigma$ relation, since
at the time of the merger any information that we can gather on the host
(via electromagnetic observations) will not provide details on the two original galaxies.  
For instance the mass ratio of the merging MBHs encodes how efficiently minor
mergers can deliver MBHs to the centre of a galaxy in order to form a
bound binary.

\begin{figure}   
  \includegraphics[width=\columnwidth]{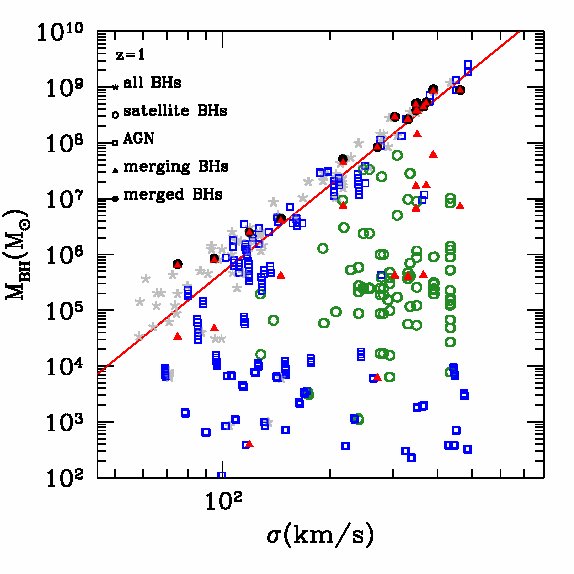}
  \caption{Dissecting the MBH population at $z = 1$: MBH population in
  our 20 ET haloes at $z=1$ (integrating over 7 time-steps, for a
  total of 0.2 Gyr). Here all MBHs evolve from the massive seeding
  model of Lodato \& Natarajan (2006), with $Q_c=2$. Stars: all
  nuclear MBHs. Empty circles: satellite/wandering MBHs. Squares:
  AGNs. Triangles: merging MBHs. Solid circles: merger remnants.  AGN
  and merging MBHS represent the detectable systems. Note that
  accreting MBHs (powering AGNs) grow notably in mass during the 7
  time-steps and progress toward the local $\msigma$ relation.}
   \label{fig5}
\end{figure}

\subsection{Hidden black holes}

Our key finding is the prediction of the existence of a large
population of hidden (as in undetectable as AGN or as merging BHs via
gravitational radiation) MBHs at all redshifts. There are two main contributors to the population of
hidden MBHs: MBHs in the \textit{nuclei} of low-mass galaxies
($\sigma\sim 20-50 {\rm km\, s^{-1}}$), and satellite/wandering MBHs.
`Hidden' nuclear MBHs have not experienced appreciable growth in mass
and formed in low mass haloes with quiet merging histories. A
potential observational signature of the `heavy seed' scenario is the
existence of a `plume' of overmassive MBHs in the \textit{nuclei} of
haloes with $\sigma\sim 20-50\, {\rm km\, s^{-1}}$.  The only way to
detect MBHs in the plume would be as IMRI/EMRI (intermediate or
extreme mass ratio inspiral) events, or via measurement of stellar
velocity dispersions and modelling as in the local universe (for
example Magorrian et al. 1998).  Approaching $z = 0$, the under-fed
part of this population likely merges into more massive galaxies.
 
Satellite and wandering MBHs would instead be off-centre systems, orbiting
in the potential of comparatively massive hosts.  Semantically we
distinguish here between MBHs that are infalling into a galaxy for the
very first time, following a galaxy merger (satellite MBHs) and those
that are merely displaced from the center due to gravitational recoil
(wandering MBHs).  Some of the {\it satellite} MBHs will merge with
the central MBH in the primary galaxy, and such merging does not
significantly alter the position of the already massive primary hole
which sits on the $M_{\rm BH} - \sigma$ relation to start with.

The MBH population in our series of simulations of the massive ET halo
is shown in Fig.~\ref{fig5}, for $z=1$. Here we dissect the MBH
population into its components.  Satellite/wandering MBHs are found
{\it below} the $\msigma$ correlation as expected (shown as open
circles, at the $\sigma$ of the host halo). Luminous AGNs are
preferentially found on the $\msigma$ relation (squares). We note the
existence of a sub-population of satellite AGNs, that is, satellite
MBHs which are actively accreting. For every pair of coalescing MBHs
(triangles), one typically sits on the $\msigma$ relation, while the
companion tends to be less massive, hence, when they merge, the
remnant finds itself in the right spot on the $\msigma$ relation
(solid circles).

\section{Conclusions and Discussion}

In this paper, we have investigated how the $\msigma$ relation is
populated at the earliest times for models with physically motivated
initial black hole seeds.  Starting with ab-initio MBH seed mass
functions computed in the context of `heavy seeds' (direct formation
of central objects from the collapse of pre-galactic discs in high
redshift haloes), or `light seeds' (Population III remnants) we follow
the assembly history to late times using a Monte-Carlo merger tree
approach. \textit{ The initial seeding does not set up the $M_{\rm BH}
- \sigma$ relation.}  In our calculation of the evolution and build-up
of mass we assume a simple prescription for determining the precise
mass gain by the MBH during a merger. Motivated by the
phenomenological scaling of $M_{\rm BH} \propto \sigma^{4-5}$, we assume
that this proportionality carries over to the gas mass accreted in
each step. This simple assumption allows us to meet a number of
observational constraints, including the luminosity function of
quasars and the mass density in MBHs at $z=0$ (Volonteri et al. 2008).

Here follows a summary of our results. 
\begin{itemize}
\item We find that the $M_{\rm BH} - \sigma$ relation can be established
early due to accretion episodes associated with major mergers even
though the original MBH seeds themselves do not satisfy this
relation. 
\item At the high mass end ($M_{\rm BH} > 10^6 \msun$), the
relation is anchored early, and low mass MBHs slowly migrate onto it
as hierarchical merging proceeds. 
\item Among active accretors, the most
massive MBHs ($M_{\rm BH} > 10^6 \msun$) sit on or around the
$\msigma$ relation at all epochs and consequently flux limited AGN
surveys are biased to preferentially detect this
population. 
\item Similarly, we find that LISA is also likely to be biased
toward detecting black holes that preferentially inhabit the $M_{\rm
BH} -\sigma$ relation. This bias is due to major mergers being more
common at high redshift for the most massive, biased, galaxies.
\end{itemize}

{\it Since we assume a priori that the accreted mass during a major
merger event scales as the fifth power of the velocity dispersion, we
inevitably recover the observed $z = 0$ slope}. Our current formalism
therefore does not equip us strictly speaking to study the evolution
of the relation or the scatter with redshift. However, the exact
scaling of the accreted mass does not affect our results, as long as
accretion is merger-driven and it establishes a clear correlation
between hole and host. To push this scenario further, we have
implemented a model where the $\msigma$ correlation evolves with
redshift as proposed by Woo et al. (2008), based on observations of
$z\sim0.5$ AGN. We have simply assigned to MBHs hosted
in galaxies experiencing a major merger a mass corresponding to the
extrapolation at all redshifts of the scaling suggested by Woo et al.: at fixed 
velocity dispersion $\log M_{BH}(z)-\log M_{BH}(0)=3.1\log (1+z) + 0.05$. This is the final mass
that a MBH would have at the end of the accretion episode.
This relation has been proposed for $z\sim0.5$  objects. We applied the same scaling all the
way to high redshift, further imposing that the MBH mass is not larger than the galaxy mass.
Implementing such rapid evolution we find overproduces the
local MBH mass density and overestimates the luminosity function of
quasars, while the main conclusions of the present paper are otherwise
unchanged.

As a further check of our result that the establishment of the $M_{\rm BH} - \sigma$  
is a function of the halo bias and hierarchy,  we have tested a model where {\it the accreted mass
does not correlate with the velocity dispersion at all.} For this
scenario, we assume a prescription for black hole growth, simply that
MBHs {\it double} in mass at every major merger with no implemented
self-regulation prescription. This model allows us to explore the
effect of the number of major mergers on MBH growth (i.e., the
connection with the cosmic bias). Although the resulting $\msigma$ has
a larger scatter at all redshifts and the local MBH mass density and
luminosity function of quasars are overestimated; we still recover a
correlation between $M_{\rm BH}$ and $\sigma$, in the sense that the
most massive galaxies do tend to host the most massive holes. Since in
this case there is no correlation between accreted mass and halo
properties, this exercise confirms that the existence of this
correlation is a pure reflection of the merger history: the most massive 
halos experience a larger number of major mergers over their lifetime,
hence their MBHs are the first to grow, and become the largest. 
The slope of  the $\msigma$ correlation is however much flatter than
the local empirical correlation ranging from 2 (for massive seeds) to
3.4 (for Population III seeds) instead of $4\div5$. We note here that
the scatter obtained in the $\msigma$ at $z = 0$ in all the models
studied here reflects both the seeding mechanism (the spread in seed
masses) and the prescription used for self-regulation.

One of our key predictions is the existence of a large population of
low mass `hidden' MBHs at high redshift which are undetectable by flux
limited AGN surveys and at merger by LISA, that at later times likely
end up as wandering MBHs. This population of low mass black holes are outliers at all
epochs on the $M_{\rm BH} - \sigma$ relation. Outliers can be detected
as IMRI/EMRI gravitational waves events, or via stellar dynamical
$M_{\rm BH}$ measurements in low--mass galaxies.  We find that nuclear
MBHs with masses in excess of $M_{\rm BH} \sim 10^6\,M_{\odot}$
preferentially lie on the $\msigma$ correlation. More accurate
measurements of MBH masses below $M_{\rm BH} \sim 10^6\,M_{\odot}$
will enable us to use the measured $z = 0$ relation to constrain
seeding models at high redshift since cosmic evolution does not appear
to smear out this imprint of the initial conditions. The scatter in
the observed $\msigma$ relation might also provide insights into the
initial seeding mechanism.  Since Population III remnants do not
appear to be efficient seeds (Alvarez et al. 2009), other channels
like the one proposed by Lodato \& Natarajan, for instance, are clearly
needed to make massive seeds. It appears that the local relation might
indeed hold clues to initial seeding mechanism.

\section*{Acknowledgments}




PN would like to thank the Radcliffe Institute for Advanced Study and the Center for
Astrophysics (CfA) for providing an intellectually stimulating atmosphere that 
enabled this work. Support for this work was provided by NASA grant NNX07AH22G 
and SAO Awards SAO-G07-8138 C and TM9-0006X  (M.V.).

\bibliographystyle{mn2e} 

\end{document}